\documentclass{eas}
\usepackage{graphicx,amsmath,amssymb}
\begin{document}

%%-----------------------------
%%      the top matter
%%-----------------------------
\title{Non-LTE Model Atom Construction}
\runningtitle{N. Przybilla: Non-LTE Model Atom Construction}
\author{Norbert Przybilla}\address{Dr. Remeis-Sternwarte Bamberg \& ECAP, Astronomisches
Institut der Universit\"at Erlangen-N\"urn\-berg, Sternwartstrasse 7, D-96049 Bamberg, Germany}
\begin{abstract}
Model atoms are an integral part in the solution of non-LTE problems. They
comprise the atomic input data that are used to specify the statistical
equilibrium equations and the opacities and emissivities of radiative transfer. A
realistic implementation of the structure and the processes governing the
quantum-mechanical system of an atom is decisive for the successful 
modelling of observed spectra. We provide guidelines and suggestions for the
construction of robust and comprehensive model atoms as required in
non-LTE line-formation computations for stellar atmospheres. Emphasis is
given on the use of standard stars for testing model atoms under a wide
range of plasma conditions. 
\end{abstract}
\maketitle
%%-----------------------------
%%      your text
%%-----------------------------
\section{Introduction}

Astrophysical plasmas like stellar atmospheres, gaseous nebulae or  accretion
disks are not in any sense closed systems, as they emit photons into
interstellar space. Therefore, the thermodynamic state of such plasmas is in
general not described well by the equilibrium relations of statistical mechanics
and thermodynamics for local values of temperature and density, i.e. by local
thermodynamic equilibrium (LTE). The presence of an intense radiation field,
which in character is very different from the equilibrium Planck distribution,
results in deviations from LTE (non-LTE) because of strong interactions between
photons and particles. The thermodynamic state is then determined by the
principle of statistical equilibrium. All microscopic processes that produce
transitions from one atomic state to  another need to be considered in detail
via the rate equations. A fundamental complication is that the distribution of
the particles over all available energy states -- the level populations or
occupation numbers -- in turn affect the radiation field via the effects of
absorptivity and emissivity on the radiation transport. What is required is a
self-consistent simultaneous solution of the radiative transfer and statistical
equilibrium equations.

A {\it model atom} is a collections of atomic input data required for the
numerical solution of a given non-LTE problem. It is a  {\it
mathematical-physical approximation} to the  quantum-mechanical system of a real
atom, and its interaction with radiation and with other particles in a plasma. A
model atom comprises, on one hand, data to specify the structure of the atom/ion
like energy levels, statistical weights and ionization potentials. On the other
hand, the transitions among the individual states need to be described,
requiring oscillator strengths, cross-sections for  photoionization and
collisional excitation/ionization, etc. The number of levels in a model atom
amounts typically to several tens to several hundred in modern work, and the
number of transitions from hundreds to many (ten-)thousands.

As only a very limited amount of atomic data have been determined experimentally
up to now -- mostly energy levels, wavelengths and oscillator strengths --, 
most of the data have to be provided by theory. Large collaborative efforts
have been made to compute the data required in astrophysical
applications via {\it ab-initio} methods. The Opacity Project (OP;
Seaton~\cite{seaton87}; Seaton et al.~\cite{seaton94}) and its successor the
IRON Project (IP; Hummer et al.~\cite{hummer93}) provided enormous databases
of transition probabilities and cross-sections for photoionization
and excitation via electron impact. Many smaller groups and individuals 
have contributed additional data,
most notably Kurucz~(see e.g.~Kurucz~\cite{kurucz06}) in a tremendous effort lasting
already for about three decades. {\it Ab-initio} data for radiative processes 
between levels of principal quantum number $n \le 10$
are available for most of the ions of the lighter elements up to calcium,
and for iron. {\it Ab-initio} data for excitation via electron collisions are by far 
less complete, typically covering transitions up to $n \le 3$~or~4 for
selected ions of the lighter elements and for iron. Reliable data for other 
members of the iron group and for the heavier elements are only selectively
available. The remainder of data -- still the bulk by number -- has to be 
approximated for practical applications.

Consequently, the starting point for the construction of model atoms for 
many elements of astrophysical interest has improved tremendously since the 
mid-1980s. Nevertheless, building realistic model atoms is neither an easy
nor a straightforward task. It is a common misconception that non-LTE 
{\it per se} brings improvements over LTE modelling. A careful 
LTE analysis of well-selected lines {\it can} be more reliable than a non-LTE
study of the `wrong' lines with an inadequate model atom. On the other hand, 
computations using a realistic model atom {\it will} improve over LTE
-- provided that the other ingredients of the modelling are also~realistic. 

The independence of microscopic processes from environment -- at least under
not too extreme conditions -- provides a tool to assess the quality of model 
atoms by comparison with observation. Comprehensiveness and robustness
of a model atom are given when it reproduces the observed line spectra over
a wide range of plasma conditions. Standard stars should serve as test
`laboratories', covering a wide range of effective temperature and surface
gravity (particle density).

In the following we discuss practical aspects of the construction of model atoms
for non-LTE line-formation calculations of trace elements  in stellar
atmospheres.  Guidelines and suggestions are given how to build up robust and
comprehensive  model atoms, and how to test them thoroughly.

\section{Problem Definition}
Non-LTE line-formation calculations solve the coupled statistical
equilibrium and radiative transfer equations for a prescribed model
atmosphere, which may itself be in LTE or non-LTE. The computational
expenses are therefore lower than for full non-LTE calculations
that solve also for the atmospheric structure. 
Hence, comprehensive model atoms may be treated in great detail,
which turns out to be a crucial advantage of restricted non-LTE calculations.

One of the most important quantities for the comparison with observed
spectral lines -- which is the basis of quantitative spectroscopy -- are
{\it accurate occupation numbers} $n_i$ for the levels involved in the
transitions. These have to be determined in general by solution of the 
rate equations of statistical equilibrium (though detailed equilibrium may turn 
out a valid approximation {\it a posteriori})
\begin{equation}
\sum\limits_{j {\neq} i} n_i (R_{ij}+C_{ij})=\sum\limits_{j\neq i} n_j
(R_{ji}+C_{ji})\,,
\label{rateeq}
\end{equation}
where the $R_{ij}$ and $C_{ij}$ are the radiative and collisional rates,
respectively, for the transitions from level~$i$~to~level~$j$. Radiative upward 
rates are given by 
\begin{equation}
R_{ij}=4\pi \int \sigma_{ij}(\nu) \frac{J_{\nu}}{h\nu}\,{\rm d}\nu\,,
\label{radrates}
\end{equation}
where $\sigma_{ij}(\nu)$ is an atomic cross-section for bound-bound or
bound-free processes, $J_{\nu}$ the mean intensity, $h$ Planck's constant
and $\nu$ the frequency. In the case of collisional processes the upward-rates 
are given by
\begin{equation}
C_{ij}=n_e \int \sigma_{ij} (v) f(v) v\,{\rm d}v\,,
\label{colrates}
\end{equation}
where $n_e$ is the electron density (for the moment we assume that
collisions with heavy particles can be neglected, see Sect.~\ref{collisions}), 
$v$ the velocity and $f(v)$ the (in general Maxwellian) velocity distribution of the 
colliding particles.
The corresponding downward rates are derived from detailed-balancing
arguments, requiring correction for stimulated emission in
the case of the radiative downward rates.

Inspection of Eqns.~\ref{rateeq}--\ref{colrates} shows which input quantities
need to be reliably known in order to facilitate an accurate determination of
the level populations:
{\sc i}) the local temperatures  and particle densities, {\it and}
{\sc ii}) the non-local radiation field, {\it and}
{\sc iii}) accurate cross-sections $\sigma_{ij}$, {\it and}
{\sc iv}) all transitions relevant for the problem have to be taken into
account. Any shortcomings in these will affect the final accuracy that can be obtained in
the modelling. 

Items {\sc i}) and {\sc ii}) depend on the prescribed model
atmosphere used for the restricted non-LTE calculations. This has to give a
fair description of the {\it real} temperature gradient and the density
stratification in the star's atmosphere under investigation. Particular care
has to be invested in the stellar parameter determination (effective
temperature $T_{\rm eff}$, surface gravity $\log g$). This has to account for a
self-consistent treatment of quantities which are often thought to be of
secondary nature (microturbulence $\xi$, metallicity, helium
abundance, etc.), see Nieva \& Przybilla (this volume).
Good knowledge of the model atmosphere code is a prerequisite for successful
non-LTE line-formation computations for individual cases -- use of
published model grids allows only to scratch the surface of the problem. 

Items {\sc iii}) and {\sc iv}) are related to the model atom. Nowadays, the
question is often not {\it whether} to use {\it ab-initio} data for the model
atom construction, but {\it which} of the available datasets to adopt. This is a
matter of experience and familiarity with atomic physics. The quality of
agreement  of {\it ab-initio} results with available experimental data should 
certainly guide the decision. A first important step is to check how well the
{\it ab-initio} calculations reproduce the observed energies of the levels in an
atom/ion. A comparison of observed and computed oscillator strengths and
cross-sections for reactions gives further indications (see for example reviews
by Williams~\cite{williams99}; Kjeldsen~\cite{kjeldsen06}). The latter will
typically  be possible only for ground and metastable states, but the
opportunity for  constraining the accuracy of the theoretical data should not be
missed. Another criterion is the agreement between length and velocity forms of
oscillator strengths to verify the internal consistency of the {\it ab-initio}
calculations.  Eventually, different model atoms may be constructed using the
alternatives as input data to decide empirically which dataset should be
preferred for the practical work.

Experience is also the key in deciding how extended a model atom should be
and which transitions should be considered. Nature realises all
possibilities, but we have to handle the mathematical solution of a set of
equations describing the physics packed into a (restricted) model. In
particular, the
numerical solution of the set of linear equations (\ref{rateeq}) requires to
be handled carefully. Precautions need to be taken to keep the numerical
problem well-conditioned and the algorithms stable. In practice this means that 
some transitions may better be ignored in a model atom, or the number of levels 
may be restricted in order to obtain meaningful results. Larger model
atoms accounting for more transitions are therefore not necessarily `better', 
even if the individual atomic input data are of high quality.

Finally, execution times of the model computations are also important for  the
practical work. They should not be excessive, requiring a certain compactness of
model atoms even for non-LTE line-formation computations on prescribed model
atmospheres.  Hence, the construction of model atoms is effectively a highly
complex optimisation problem. A compromise needs to be found between
comprehensiveness, accuracy, stability and efficiency.

\section{Model Atom Structure}\label{structure}

The first step in the construction of a model atom is to decide on the
extension  of the model: which ions, which energy levels should be included, and
how? This depends on the specific non-LTE problem and may range from a few
levels for studies of a resonance line to many hundred -- including packed
`superlevels' --  if a reproduction of the complex spectra of e.g. iron group
species is aspired.  In the following we concentrate on the general strategy for
the construction of comprehensive model atoms, which are able to reproduce
practically the entire observed spectra of an ion over a wide parameter range. 

\begin{figure}[t]
\begin{center}
\includegraphics[width=.65\linewidth]{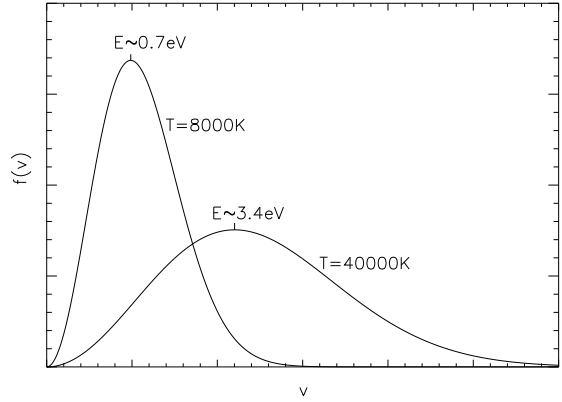}
\end{center}
\vspace{-7mm}
\caption{Schematic plot of the Maxwellian velocity distribution for typical
temperatures at line-formation depths in A- and O-type stars. 
}
\label{maxwell}
\end{figure}

\begin{figure}[t]
\begin{center}
\includegraphics[width=.9\linewidth]{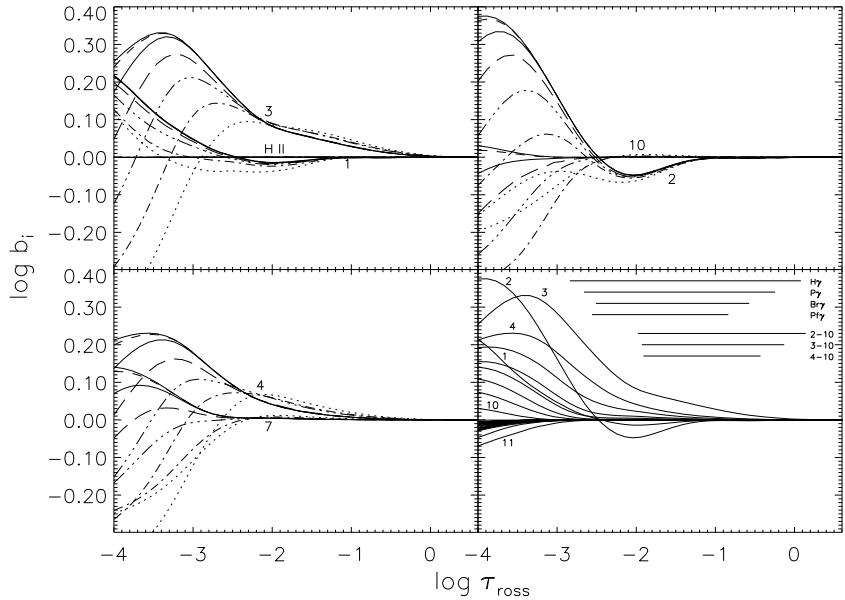}
\end{center}
\vspace{-7mm}
\caption{
Run of departure coefficients $b_i$ in a model of the bright supergiant
$\beta$\,Ori (B8\,Ia) as a function of $\tau_\mathrm{ross}$ for H\,{\sc i}
model atoms of different complexity: for 10,
15, 20, 25, 30, 40, 50 levels (dotted, dashed-dotted,
dash-dot-dot-dotted, long dashed, full, dashed, full thick lines). 
The lower right panel displays the behaviour of all levels in the 50-level
model atom. Note the Rydberg states asymptotically approaching the H\,{\sc ii} 
limit. See the text for details. From Przybilla \& Butler~(\cite{PrBu04}).
}
\label{departureconvergence}
\end{figure}

\clearpage

The fundamental parameter that determines which ions should be included in a
model atom is the effective temperature, as this determines the energetics of
the  microscopic processes. The term structure of the ions provides a second
criterion. Hence, the main ionization stage should be adequately
represented, plus usually two or three minor ionic species --  which may
comprise in fact the ones of particular interest. E.g., the main ionization
stage of carbon in a B0\,V star (with $T_{\rm eff} \approx 32\,000$\,K) is
C\,{\sc iii} in the line-formation region, but features of the minor species
C\,{\sc ii} and C\,{\sc iv}  are also present in the optical spectra.
Consequently, a comprehensive model atom should consider C\,{\sc ii-iv}, plus
the ground level of C\,{\sc v}. The latter has an enormous energy gap between
the ground and the first excited level, of about 300\,eV, such that from atomic
structure considerations excited C\,{\sc v} levels can be safely ignored in the
model atom. A lithium model atom for use with solar-type stars can be kept
simple  -- detailed Li\,{\sc i} + the Li\,{\sc ii} ground level -- for the
same reason, despite Li\,{\sc ii} dominating by far in terms of population.

Concerning the choice of levels to include in a comprehensive model atom
there are two objective criteria available, which are tightly coupled. One is
the energy gap between the energetically highest level of the ion and the 
continuum, as defined by the ground level of the next higher ionization stage. 
As a general rule the gap should be less than $kT$ to ensure an accurate
determination of ionization fractions\footnote{Minority
species are particularly sensitive to non-LTE effects because any
small change of the ionization rates can affect their populations by a
significant amount.}. In order to
keep a model atom robust it is recommended to include energetically higher
levels than the minimum necessary to cope with a given problem. Non-LTE
studies of the Mg\,{\sc ii} $\lambda$4481\,{\AA} line may serve as an example: the
line is observable from F-type to mid O-type stars. For explaining the
behaviour of the line in early-type stars (Mihalas~\cite{Mihalas72}) it may be
sufficient to consider levels up to $n$\,=\,5 or 6 using the criterion
above, as the majority of the colliding particles will be able to facilitate
ionization (see Fig.~\ref{maxwell}). However, the same model atom will not be
useful for analyses of A-type stars, as only a small fraction of electrons in the
high-velocity tail of the Maxwell distribution will be able to overcome the
reaction threshold. In that case, completeness up to $n$\,=\,8 would be required, etc.

\begin{figure}[t]
\includegraphics[width=.49\linewidth]{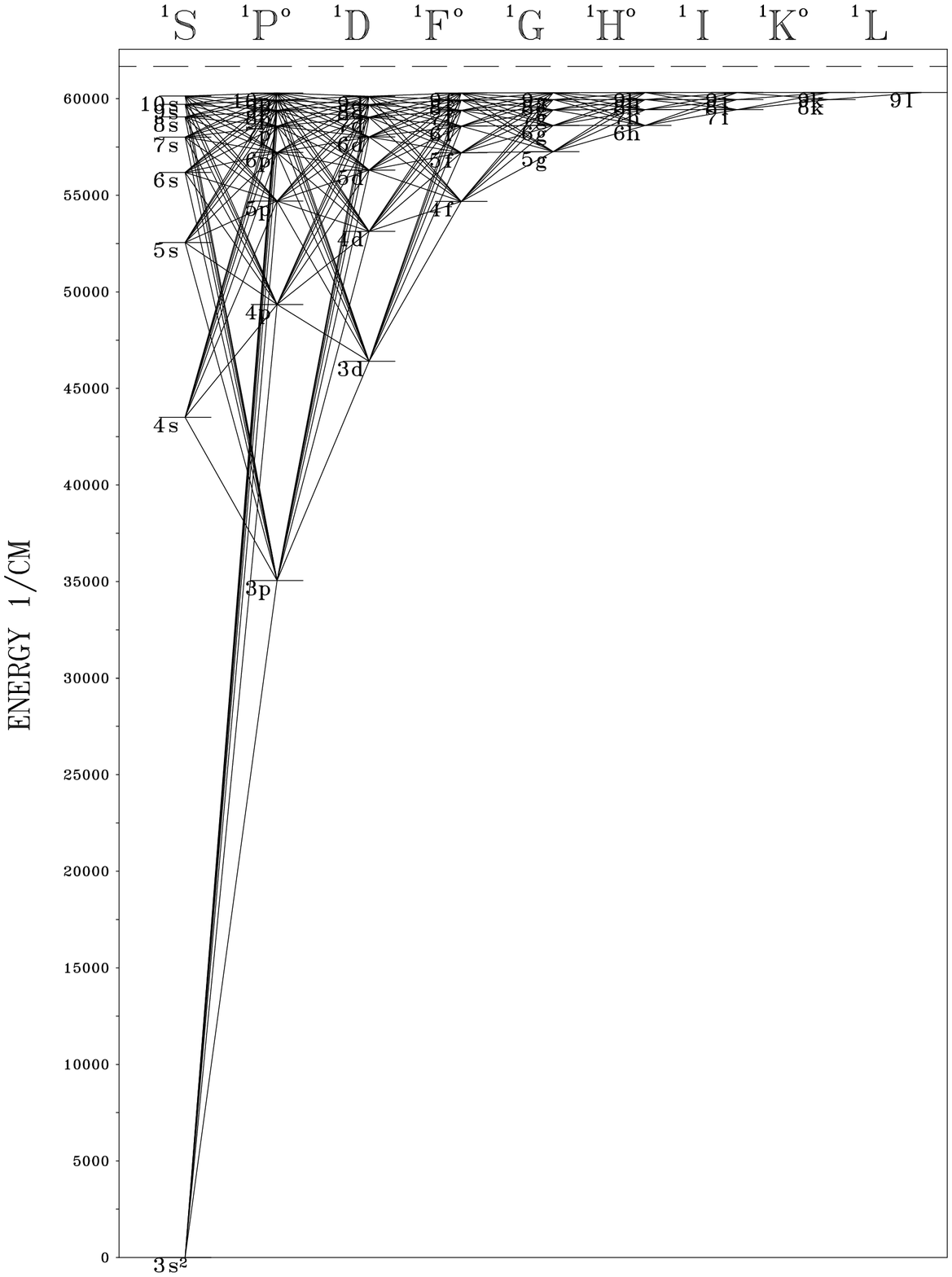}
\hfill
\includegraphics[width=.49\linewidth]{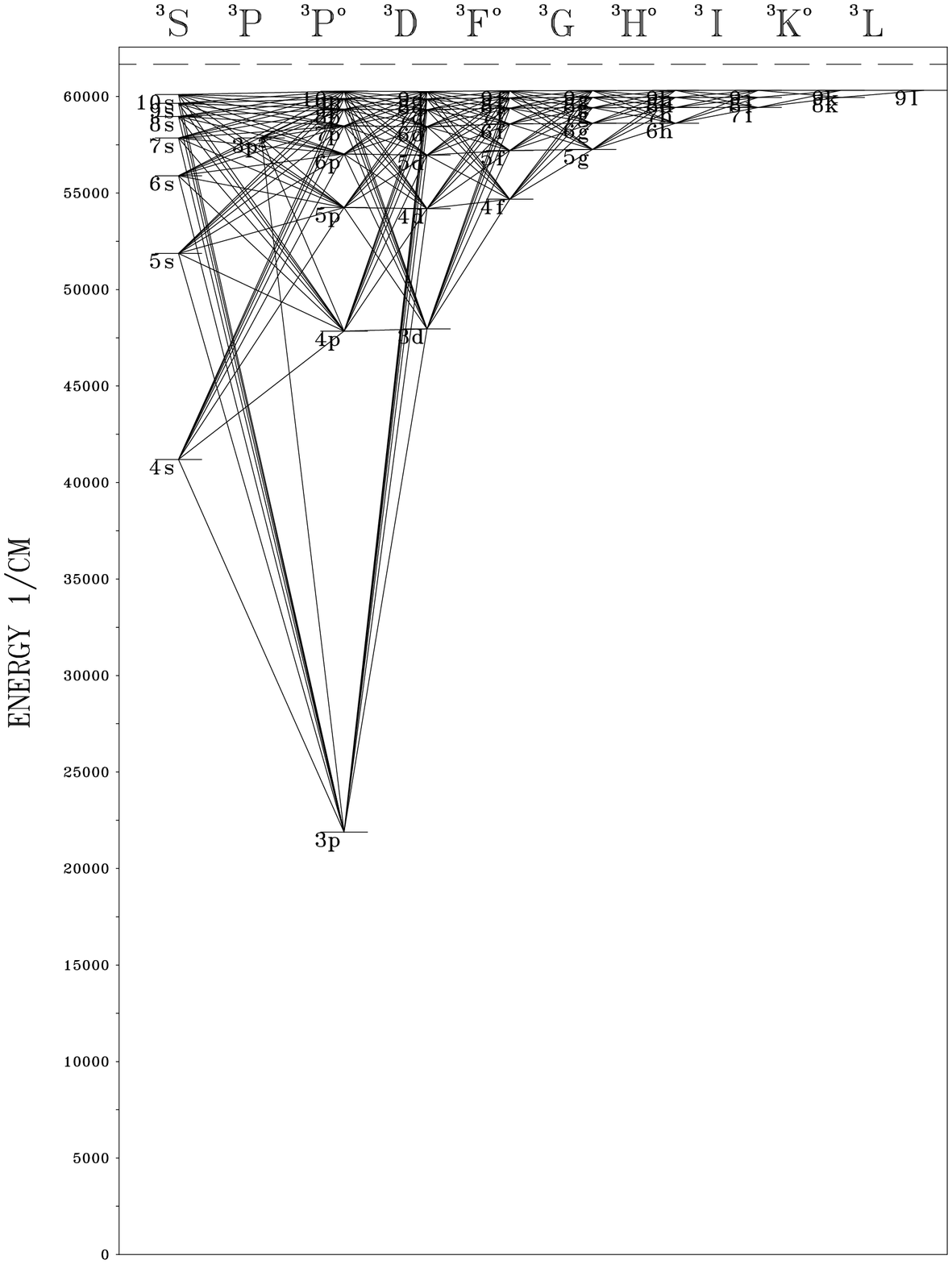}
\vspace{-5mm}
\caption{
Example of a comprehensive model atom structure. Grotrian diagrams for the
singlet and
triplet spin systems of Mg\,{\sc i} are shown. Displayed are the radiative bound-bound
transitions treated explicitly in the non-LTE calculations. The ionization threshold
is in\-dicated by the dashed line. From Przybilla et
al.~(\cite{Przybillaetal01}).
}
\label{grotrian}
\end{figure}

The second criterion is the convergence of the behaviour of level departure 
coefficients $b_i$ with increasing model complexity (for a given set of
transition data). An example is shown in Fig.~\ref{departureconvergence}, for
hydrogen model atoms considering levels up to $n$\,=\,10 to 50.  Models with a
too low a number of levels can show a different behaviour at line-formation
depths than the more complex models, resulting in inaccurate predictions. An
alternative formulation of this criterion is via the convergence of the line
source function (e.g. Sigut~\cite{sigut96}).  

For many elements fine-structure states of a term may safely be combined  into
one level representing the term. This comprises, in particular, the cases  that
are approximated well by $LS$-coupling. Collisions couple the individual
sub-levels tightly because of their small energy separations, i.e. they are in
LTE relative to each other. A similar opportunity for simplification of the
model atom opens up for levels at higher excitation energies. The energy
separations of terms decrease with increasing angular quantum number  $\ell$ for
the same $n$, and in general with increasing $n$. Eventually, they may be safely
grouped into one level with appropriate statistical weight. This helps to keep
the number of explicit non-LTE levels to be treated for elements up to about
calcium below $\sim$200, even if several ionization stages are treated
simultaneously. An example of a comprehensive model atom structure is visualised
in Fig.~\ref{grotrian} in form of a Grotrian diagram, for Mg\,{\sc i}.  For the
heavier elements with complex electron structure like the iron group  elements
it may be worthwhile to consider regrouping a multitude of levels with similar
properties into  `superlevels' (and the transitions into `superlines'), a
concept first introduced by Anderson~(\cite{Anderson89}).

\section{Radiative Transitions}\label{radiation}
The {\em non-local} character of the radiation field drives the stellar
atmospheric plasma {\em out of detailed equilibrium}: 
photons can travel large distances before interacting with the particles, 
coupling the thermodynamic state of the plasma at different depths in
the atmosphere. This affects the excitation and ionization of the
material. Radiative transitions obey {\em selection rules}.

\subsection{Line Transitions}
Changes in the internal energetic state of an atom/ion can occur by
the absorption/emission of photons, giving rise to spectral lines.
The strength of a spectral line is basically determined by the number of
absorbers (or emitters) and the line absorption cross-section, which is given by
%(unaccounted for stimulated emission)
\begin{equation}
\sigma_{ij}=\frac{\pi e^2}{m_ec}f_{ij}\phi(\nu)\,, 
\label{rbb}
\end{equation}
where $e$ is the electron charge, $m_e$ the electron mass and $f_{ij}$ the
oscillator strength. $\phi(\nu)$ is the line absorption profile (the
emission line profile is identical assuming complete redistribution), which
can be approximated well by a Doppler profile in most cases when the coupled
radiative transfer and statistical equations are solved to determine
the level populations\footnote{Detailed line broadening theories
become important for the last step of non-LTE line-formation
computations, when synthetic spectra are calculated.} in the
restricted non-LTE approach.

The accuracy of the oscillation strengths used for the  model atom construction
will be a limiting factor for analyses.  Fortunately, high-accuracy data is
available in many  cases\footnote{Oscillator strengths are the only ones among
the atomic data discussed further on in Sects.~\ref{radiation}--\ref{other} that
are required in LTE investigations.}, obtained both from experiments as well as
from {\em ab-initio} calculations. 

\begin{figure}[t]
\begin{center}
\includegraphics[width=.65\linewidth]{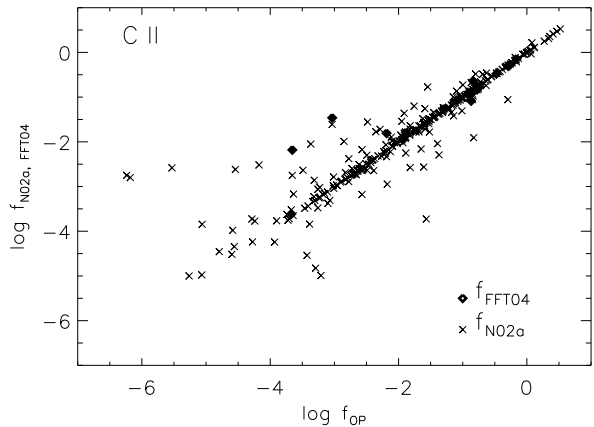}
\end{center}
\vspace{-5mm}
\caption{
Comparison of oscillator strengths for C\,{\sc ii}: values from
Froese Fischer \& Tachiev~(\cite{FFT04}) and Nahar~(\cite{Nahar02})
vs. data from the Opacity Project (Yan et al.~\cite{Yanetal87}). From
Nieva \& Przybilla~(\cite{NiPr08}, NP08).}
\label{rbbplot}
\end{figure}

It is worthwhile to cross-check available data sources. Newer data 
are not necessarily better, even if a apparently more advanced method
was used for their determination. As usual, the devil is in the
details. An example is shown in Fig.~\ref{rbbplot}. Oscillator strengths 
for C\,{\sc ii} from the OP ($R$-matrix method in the close-coupling approximation
assuming $LS$-coupling) are compared to data of Nahar~(\cite{Nahar02}) obtained with the (in
principle superior) Breit-Pauli $R$-matrix (BPRM) method  and the smaller
number of high-precision data from application of the Multiconfiguration 
Hartree-Fock (MCHF) method. The in general good agreement between the
OP and the MCHF data (with few exceptions they follow the 1:1
relation) and the considerable scatter in the comparison of OP and 
Nahar's $f$-values was one of the factors to disregard this particular set of
BPRM data from the model atom construction in that case.
However, the question which of the available data should be used for the
construction of model atoms has no simple answer in general. 

Limitations on the number of transitions that can be considered
explicitly in the radiative transfer calculations may be imposed by
the numerical method chosen for solving the equation systems. Complete
linearisation techniques are much more restricted than Accelerated
Lambda Iteration (ALI) techniques. The most important transitions 
(high $f$-value, location at wavelengths with non-negligible flux)
will need to be considered in the former case while virtually all 
line transitions may be accounted for in the latter case, see
Fig.~\ref{grotrian} for an example.

\subsection{Photoionizations}\label{photoionizations}
\begin{figure}[t!]
\begin{center}
\includegraphics[width=0.99\linewidth]{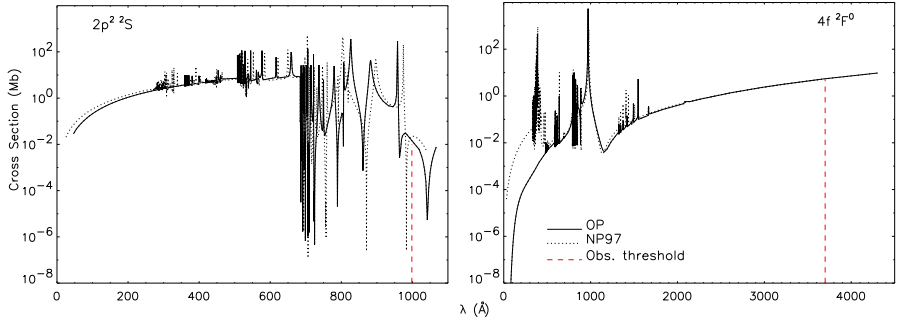}
\end{center}
\vspace{-6mm}
\caption{Comparison of C\,{\sc ii} photoionization cross-sections from 
the Opacity Project (Yan \& Seaton~\cite{YaSe87}) and Nahar \& Pradhan~(\cite{NP97}) for
2$p^2$\,$^2$S and 4$f$\,$^2$F$^{\rm o}$ as a function of wavelength.
From~NP08.}
\label{photocross}
\end{figure}

The availability of photoionization cross-sections has improved
enormously in the past 20 years, mostly because of the efforts made in
the OP and IP. Like for oscillator strengths, the quality of the different data sets
should be evaluated for the model atom construction. A comparison of
photoionization cross-sections from two different calculations is
shown in Fig.~\ref{photocross}, for two excited levels in C\,{\sc ii}.
Maybe surprisingly, some details can be relevant for the model atom construction, 
while obvious discrepancies may be not. The small shifts in the
resonance structure near the ionization threshold of the energetically
relatively low-lying 2$p^2$\,$^2$S state of C\,{\sc ii} (see
Fig.~\ref{testing} for a Grotrian diagram) can be important
for the rate determination, as they occur near the flux maximum of
OB-type stars. The more than two orders-of-magnitude differences in the high-energy 
tail of the cross-section for the 4$f$\,$^2$F$^{\rm o}$ level 
(occurring because of the different targets in the two {\em ab-initio} 
calculations) are irrelevant on the other hand, because of the low
flux at these short wavelengths. It has to be decided by comparison of
the results from different model atom realisations which of the available data 
should finally be selected, see Sect.~\ref{tests} for an example. As
always, some guidance may also be obtained by comparison of the
different theories with experiments.

The OP and IP have provided {\em ab-initio} data for photoionization
from levels with typically  $n \le 10$ and $\ell \simeq 4$. Missing data 
for levels of higher $n$ or $\ell$ may be rather safely approximated by 
hydrogenic cross-sections (Mihalas~\cite{Mihalas78}, p.99)
\begin{equation}
\sigma(\nu)=2.815\times10^{29} \frac{Z^2}{n^5} \frac{g_{\rm II}(n,\nu)}{\nu^3}\,,
\label{hydrogenic}
\end{equation}
for the model atom construction, in particular as these levels are
usually packed, see Sect.~\ref{structure}. Here,
$Z$ denotes the charge of the ion and $g_{\rm II}$ the bound-free Gaunt factor,
which is of order unity at ionization threshold. The threshold
cross-section is given by 
$\sigma (\nu_0, n) = 7.91\times10^{-18}/Z^2~n\,g_{\rm II}$~cm$^2$.

\section{Collisional Transitions}\label{collisions}
Inelastic collisions with particles can also lead to excitation and
ionization of atoms/ions. The velocity distribution of particles in stellar 
atmospheres is in practically all cases Maxwellian, determined by the {\it local}
plasma temperature. Collision processes will therefore drive the plasma
{\it towards LTE}. Contrary to radiative transitions, no selection rules apply.
Typically, only electron collisions are considered, as
the thermal velocity and hence the collision frequency of heavy particles 
is much smaller, e.g. by a factor $\sim$43 for hydrogen. This is in
particular valid for all environments where the stellar plasma is
sufficiently ionized. Heavy particle collisions may become important in 
special cases, however, like for cool metal-poor stars. 

\subsection{Collisional Excitation}
The main aspect of collisional excitation by electron impact 
is certainly the thermalising effect on level populations in general, 
dampening out departures from detailed equilibrium imposed by the radiative
processes. Curiously enough, collisional excitation can also drive
individual level occupations {\em out of LTE} under special circumstances. 
This occurs for cases where a level is collisionally tightly-coupled to
another level that experiences strong non-LTE departures. Examples are
energetically close levels from different spin systems, 
one being metastable (a radiative decay to the ground state is
prohibited by selection rules). A prominent example for this 
coupling are the 3$s$\,$^3$S and 3$s$\,$^5$S levels of O\,{\sc i}
(Przybilla et al.~\cite{Przybillaetal00}; Fabbian et
al.~\cite{Fabbianetal09}).

Larger sets of collisional excitation data for transitions up to
typically $n$\,=\,3 or 4 are available now,
e.g. from the IP. A good part of the data are published in the
astrophysics literature, but the reader should be aware that much more
data can be found in the physics literature because of their relevance
for fusion research and technical applications.
Most useful for practical applications are 
tabulations of thermally-averaged effective collision strengths
\begin{equation}
\Upsilon_{ij}=\int_0^{\infty} \Omega_{ij} \exp(-E_j/kT)\,{\rm
d}(E_j/kT)\,,
\label{upsilon}
\end{equation}
where $E_j$ is the energy of the outgoing electron and $\Omega_{ij}$ the
collision strength, which is symmetric as well as dimensionless, 
$\Omega_{ij}=\Omega_{ji}$. Effective collision strengths facilitate an
easy evaluation of transition rates, which are proportional to
$\Upsilon_{ij}$.

While {\em ab-initio} data are of highest relevance for the construction of
model atoms, one has to resort to approximation formulae  for the vast majority
of possible transitions. Different descriptions are available from the
literature, a comparison can be found in Mashonkina~(\cite{Mashonkina96}). In
the following we concentrate on the two most commonly used approximations.

\begin{figure}[t!]
\begin{center}
\includegraphics[width=0.75\linewidth]{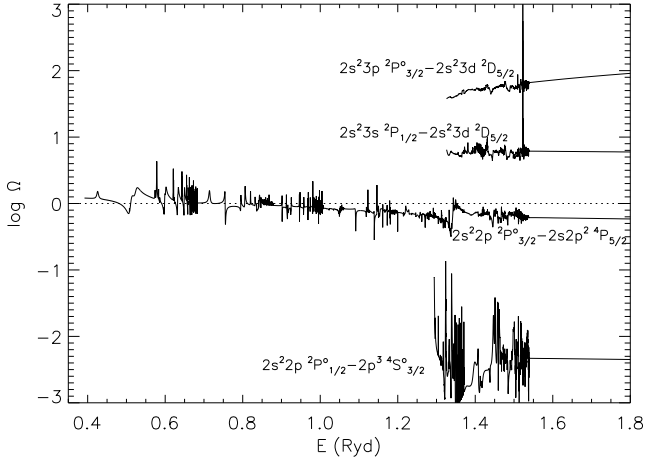}
\end{center}
\vspace{-6mm}
\caption{
Examples of collision strengths $\Omega$ for optically allowed and forbidden
transitions in C\,{\sc ii} in the near-threshold region as a function of
incident electron energy (fine-structure data from Wilson et
al.~\cite{wilson05}).
For comparison, $\Omega = 1$ is indicated by a dotted line.
}
\label{omega}
\end{figure}

The semiempirical formula of Van Regemorter~(\cite{VanRegemorter62})
allows rates for radiatively permitted transitions to be evaluated
in terms of the oscillator strength, 
\begin{equation}
C_{ij}=5.465\times10^{-11} n_e T^{1/2} [14.5 f_{ij} (u_{\rm H}^2/u_0)]
\exp(-u_0) \Gamma(u_0)\,, 
\label{vanregemorter}
\end{equation}
where $u_{\rm H}$ is the ionization
potential of hydrogen scaled by $kT$, $u_0 = E_0/kT$ ($E_0$ is the
threshold energy for the process), and $\Gamma
(u_0) \equiv {\rm max}[\bar{g}, 0.276\exp (u_0)E_1(u_0)]$ for ions 
($E_1$ is the first exponential integral). The parameter $\bar{g}$ is about 0.2 if the
principal quantum number changes during the transition and about 0.7 if 
not. For neutral atoms $\Gamma (u_0)$ has a different form (see Auer \&
Mihalas~\cite{AuMi73}).

Excitation rates of radiatively forbidden transitions are often
evaluated according to the formula of Allen~(\cite{Allen73}),
\begin{equation}
\Gamma_{ij}(T)=\frac{\Omega_{ij}}{g_i} \frac{h\nu_{\rm H}}{kT}\,,
\label{allen}
\end{equation}
where $\Gamma_{ij}(T)$ is a temperature-dependent factor in the collision
rate as defined by Mihalas~(\cite{Mihalas78}, p.~133). Typically,
$\Omega = 1$ is assumed for forbidden transitions.

Note that Eqns.~\ref{vanregemorter} and \ref{allen} give,
at best, order-of-magnitude estimates. Comparisons of {\em ab-initio}
data with the approximations should be made whenever possible to
evaluate the true uncertainties that can be expected. An example is
shown in Fig.~\ref{omega}. Differences of
up to several orders of magnitude are found, and the
quantum-mechanical data can show pronounced resonance structure.
It is therefore advisable to investigate the available {\em ab-initio}
data for trends and regularities. Extrapolations can be made based on
these in order to improve the approximate input data for the model
atom construction.

\subsection{Collisional Ionization}
The rates of collisional ionization by electron impact are mostly
affected by the availability of electrons with energies high enough to
overcome the threshold for the reaction. This makes collisional
ionization from the ground state and energetically low-lying levels
rather inefficient, as only few electrons in the high-velocity tail
of the Maxwell distribution are available for this. On the other hand,
collisions become a dominant factor for the coupling of high-lying
levels to the continuum.

Unfortunately, cross-sections for ionization by electron impact are among the
least-constrained atomic data. Experiments usually cover
ionization from the ground state only, and the reader is referred to
the atomic physics literature for extracting data for a specific problem. 
On the theoretical side, few methods have been successfully applied. 
The fundamental challenge which distinguishes collisional ionization
from excitation is the fact that the Coulomb interaction between each
of the two outgoing electrons and the residual ion is present even at
large distances. Recently, breakthrough results have been obtained by
use of the converging close-coupling method. Several fundamental processes 
have been modelled accurately, providing cross sections that closely reproduce 
the available experimental data in these cases,
see Bray et al.~(\cite{Brayetal02}) for a review. 

However, in the majority of cases one has to rely on more simple
approaches. An often used approximation formula for quantifying collisional 
ionization was given by Seaton~(\cite{Seaton62}), which expresses the collisional cross-section in
terms of the photoionization cross-section, yielding a rate
\begin{equation}
C_{i\kappa}=1.55\times10^{13} n_e T^{-1/2} \bar{g}_i \sigma(\nu_0) \exp(-u_0)/u_0\,,
\label{seaton}
\end{equation}
where $\sigma(\nu_0)$ is the threshold photoionization cross-section
(Sect.~\ref{photoionizations}), and
$\bar{g}_i$ is of the order 0.1, 0.2, and 0.3 for $Z$\,$=$\,1, 2, and
$\ge$3, respectively. Again, Eqn.~\ref{seaton}
provides, at best, an order-of-magnitude estimate. A
comparison of collision rates calculated with Eqn.~\ref{seaton} 
with those evaluated by an empirical analytical expression
(Drawin~\cite{Drawin61}) indicates that the uncertainties may be
sometimes much larger (Mashonkina~\cite{Mashonkina96}). The Seaton
formula (and analogous simple approximations) should therefore 
be applied with caution. 

\subsection{Hydrogen Collisions}
Excitation and ionization by inelastic heavy particle collisions are usually
considered unimportant in comparison to electron collisions, which
occur more frequently. 
However, the ratio of hydrogen atoms to electrons may easily
exceed a factor 10$^4$ in cool stars, in particular in metal-poor objects.
In such a case H collisions may become a dominant thermalisation
process, which must not be neglected. 

Characteristic collision energies in cool star photospheres are of the
order $kT \approx 0.2-0.6$\,eV, 
i.e. they are comparable to typical atomic transition energies.
Consequently, the near-threshold behaviour of the cross-sections
is most important for the determination of the collision rates.
Laboratory measurements or {\it ab-initio} calculations of cross-sections
near threshold are scarcely found in literature for these neutral particle
collisions. Some progress has been made 
recently for the Na~+~H system (Fleck et al.~\cite{Flecketal91}; Belyaev et
a.~\cite{Belyaevetal99}, \cite{Belyaevetal10}), showing that the
collision rates can differ by several orders of magnitude 
compared to simple approximations (see the discussion below).
Similar discrepancies were also found for the Li~+~H system
(Belyaev \& Barklem~\cite{BeBa03}). 

In face of the absence of reliable data for practically all cases of
interest, one has to resort to the use of approximations for the description of
inelastic hydrogen collisions. Most work relies on the Steenbock \&
Holweger~(\cite{StHo84}) approximation, who generalised Drawin's formulae
(Drawin~\cite{Drawin68, Drawin69}; based on Thomson's classical theory), 
originally developed for collisions between identical particles. 
The Maxwell-averaged formulae for excitation and ionization of
particle species $A$ by collision with~H~becomes
\begin{equation}
\langle\sigma v\rangle=16\pi a_0^2 \left( \frac{2kT}{\pi\mu}\right)^{1/2}Q\,\frac{m_{\rm
A}}{m_{\rm H}}\frac{m_e}{m_{\rm H}+m_e}\Psi (W)\,,
\label{sh}
\end{equation}
with $\mu = m_{\rm A}m_{\rm H}/(m_{\rm A}+m_{\rm H})$ denoting the reduced
mass. For collisional excitation $Q = (I_{\rm H}/\Delta E)^2 f_{lu}$ and
$W = \Delta E/kT$ and for collisional ionization $Q = (I_{\rm H}/I_{\rm A})^2
f_i \xi_i$ and $W = I_{\rm A}/kT$, where $\Delta E$ denotes the energy 
difference between upper and lower state of the transition and $I_{\rm H}$ and
$I_{\rm A}$ are the ionization energy of hydrogen and atomic species $A$; $f_{lu}$
is the oscillator strength, $f_{i}$ is an efficient oscillator strength for
ionization and $\xi_i$ the number of equivalent electrons. The function
$\Psi (W)$ is given by
$\displaystyle \Psi (W) = (1+\frac{2}{W}) \exp(-W) (1+\frac{2m_e}{(m_{\rm H}+m_e)W})^{-1}$.
Note that Eqn.~\ref{sh} does not apply to optically forbidden
transitions. Takeda~(\cite{Takeda91}) suggested to relate the hydrogen
collision rate with the electron collision rate via
\begin{equation}
C_{ij}^{\rm H}=C_{ij}^e\frac{n_{\rm H}}{n_e}\left( \frac{m_e}{\mu} \right)^{1/2}\,,
\label{tak}
\end{equation}
assuming {\em ad-hoc} a similarity of cross-sections for both cases
($C_{ij}^e$ is usually evaluated with $\Omega = 1$).
Here, $n_{\rm H}$ denotes the number density of neutral hydrogen.

Often the results are scaled by a factor $S_{\rm H}$. A value of $S_{\rm H} = 0$
is equivalent to no hydrogen collisions, $S_{\rm H} \rightarrow \infty$
enforces LTE. In general, $S_{\rm H}$ is constrained empirically by demanding
that the scatter in abundance as determined from the entire sample of lines of a
species should be minimised.

Equations~\ref{sh} and \ref{tak} were considered appropriate to provide an 
order-of-magnitude estimate for a long time, but in view of the few {\em ab-initio} calculations 
available now, this assessment appears too optimistic. This confirms earlier
indications of an underestimation of the real uncertainties, from the
comparison with other approximations, see e.g.
Mashonkina~(\cite{Mashonkina96}). 
Barklem~(\cite{Barklem07}) investigated the uncertainties for the most
simple case, H~+~H, in detail.

The Steenbock \& Holweger formula remains in use 
for astrophysical applications in view of the lack of other reliable
data, despite all the warning evidence. In view of this, efforts
should be made to determine proper scaling factors $S_{\rm H}$ in
order to minimise the impact on the accuracy of analyses. This could
be made by extensive investigations following the recommendations for
testing model atoms in Sect.~\ref{tests}, covering the parameter space
comprehensively (wide range of effective temperature, densities and metallicity).

\section{Other Processes}\label{other}
Consideration of the processes described in the previous two sections is usually 
sufficient for the construction of model atoms for the analysis of
stellar photospheres. Nonetheless, two other
types of radiative and collisional processes may be of relevance in
some cases. They are only briefly described in the following for
completeness, leaving it to the reader to investigate the specialist literature.

{\em Autoionization} can occur in complex atoms with several
electrons. When two electrons are excited simultaneously, this can
give rise to states below and above the ionization potential. The
states above the ionization threshold may autoionize to the ground state
of the ion, releasing one electron. The inverse process is also
possible, and, if a stabilising radiative decay occurs within the
(short) lifetime of the doubly excited state\footnote{The presence of
these so-called autoionizing states has consequences for the
absorption of photons and the scattering of electrons by atoms/ions:
it gives rise to resonances in the photoionization and electron
collision cross-sections, see Figs.~\ref{photocross} and \ref{omega}.}, 
it can give rise to an
efficient recombination mechanism. This is referred to as {\em
dielectronic recombination}. Details of rate coefficient modelling can
be found e.g. in Badnell et al.~(\cite{Badnelletal03}). An application 
in the context of WR-type central stars of
planetary nebulae is discussed by de Marco et al.~(\cite{deMarco98}).

{\em Charge exchange} reactions are collisional processes between
atoms/ions in which one, or more, electrons are transferred from one
collision partner to the other, e.g.
A\,$+$\,B$^+$\,$\rightarrow$\,A$^+$\,$+$\,B, with B usually being H or He.
One well-known reaction occurring in stellar atmospheres is
O\,$+$\,H$^+$\,$\rightleftharpoons$\,O$^+$\,$+$\,H, which can dominate
the ionization balance of oxygen as the non-LTE departures of the $n$(H\,{\sc
i})/$n$(H\,{\sc ii}) are forced upon $n$(O\,{\sc i})/$n$(O\,{\sc ii})
by this resonant reaction (the ionization potentials of neutral
hydrogen and oxygen are very similar), see e.g. Baschek et
al.~(\cite{Bascheketal77}) or Przybilla et
al.~(\cite{Przybillaetal00}) for a discussion. 
A list of reaction playing a
possible role in astrophysical plasmas and tabulated
reaction rates are provided e.g. by Arnaud \& Rothenflug~(\cite{ArRo85}).

\section{Testing Model Atoms}\label{tests}
The question whether a model atom is realistic can only be answered by
comparison with observation. One needs to test whether the model atom
is comprehensive enough, i.e. whether the level structure and all
relevant connecting transitions are considered properly and whether the atomic data
used are sufficiently accurate. Usually, this should give rise to an
iterative process: a stepwise improvement of the model atom by empirical selection of the
`best' input data. The aim is to single out {\em one} model atom, that reproduces the
observed spectra closely {\em at once}, independent of the plasma conditions 
(atomic properties are independent of environment).

\begin{figure}[t!]
\begin{center}
\includegraphics[width=0.43\linewidth]{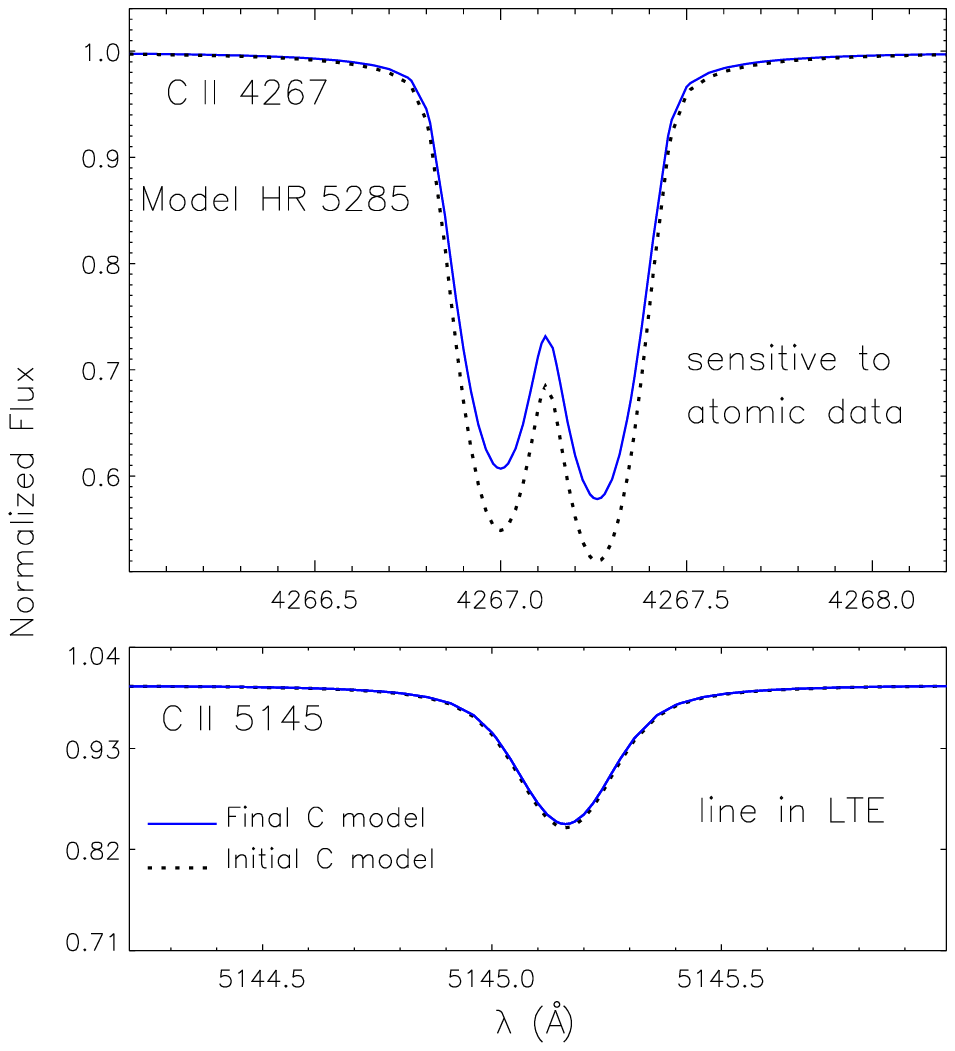}
\hfill
\includegraphics[width=0.56\linewidth,height=5cm]{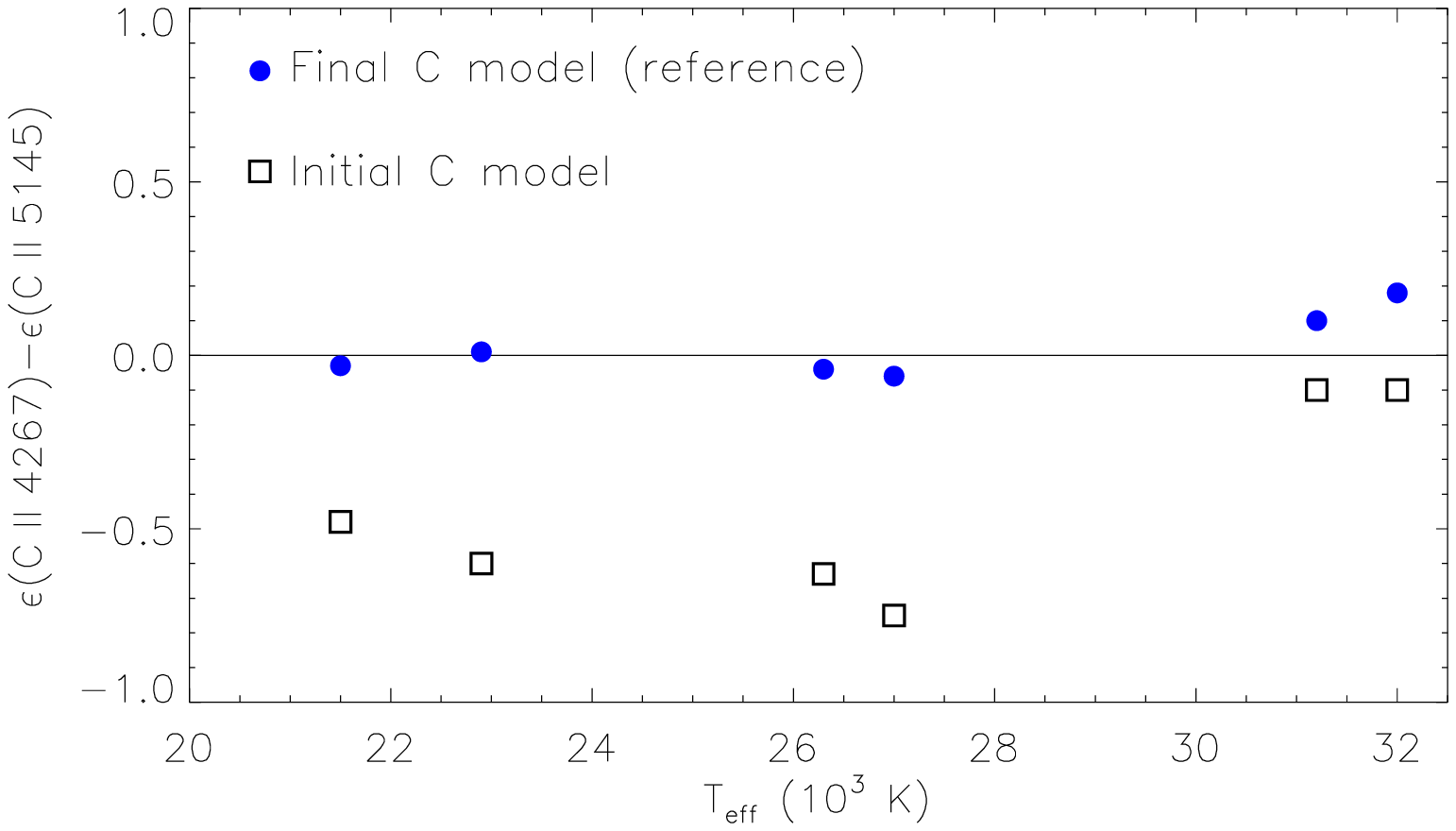}
\end{center}
\vspace{-6mm}
\caption{Left: sensitivity of two C\,{\sc ii} lines
to use of photoionization cross-sections from different {\it ab-initio}
calculations in a model spectrum for a B2\,V star.
Right: differences in (logarithmic) carbon abundance as derived from
the two lines in 6 stars as a function of effective temperature. From~NP08.}
\label{photoeffects}
\end{figure}

In order to give an idea on the practical approach for performing such
tests we discuss an example. Synthetic line profiles from
calculations with two different model atoms are compared in the left panel 
of Fig.~\ref{photoeffects}. While the strong C\,{\sc ii}
$\lambda$4267\,{\AA} transition is highly sensitive to non-LTE effects
-- in particular to the photoionization cross-sections adopted --, the
other line is virtually insensitive to any model atom realisation
using reasonable atomic input data (it is `in LTE'). Such a sensitivity
is one of the keys to select the `right' photoionization data for the model
atom construction. The second ingredient in this process is the comparison with 
observations, here for stars in a temperature sequence 
(right panel of Fig.~\ref{photoeffects}). This is in order to test the
reaction of the model realisations to a hardening radiation field.
The line `in LTE' serves as reference and the goal of the model atom
optimisation is to minimise the differences in the derived abundances
from the various indicators. In this case it was shown that ill-chosen atomic data 
can result in line abundance differing by up to 0.8\,dex
(NP08), which helped to solve one notorious non-LTE problem of
stellar astrophysics (Nieva \& Przybilla~\cite{NiPr06}).

\begin{figure}[t!]
\begin{center}
\includegraphics[width=.86\linewidth]{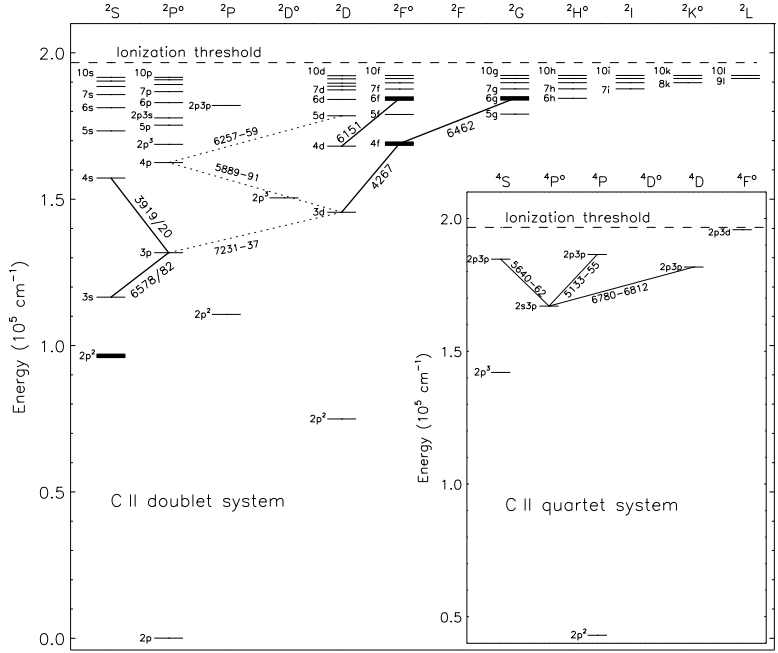}
\end{center}
\vspace{-7mm}
\caption{Grotrian diagram for the C\,{\sc ii} doublet and quartet spin
systems. Transitions that give rise to lines in the optical
spectra of B-type stars are identified by continuous and dotted lines. 
These are the observables available for testing model atoms.
Additional channels become available in the UV and IR spectral range.
From~NP08.
}
\label{testing}
\end{figure}

Ideally, the observations used for the model atom calibration should range 
from the far-UV to the near-IR 
in order to test as many channels (spectral lines) as possible, even
those that may be irrelevant for practical applications later. At the
same time, the observations should cover an as wide range of plasma
parameters as possible: high-density, i.e. collisionally
dominated, environments like the photospheres of dwarf stars
and low-density plasmas (those dominated by radiative
processes) as encountered in (super-)giants should be considered, at different
temperatures. Where possible, also the metallicity dependence of
non-LTE effects should be investigated to test the response of a
model atom to a varying radiation field, e.g. by considering
stars of Population~I and II. Further tests may include
more `exotic' environments, like He-dominated plasmas in compact
subdwarfs (Przybilla et al.~\cite{Przybillaetal06a}) or
giant extreme helium stars (Przybilla et 
al.~\cite{Przybillaetal05},~\cite{Przybillaetal06b}).

Such a comprehensive approach involving satellite observations 
may not be feasible in almost all cases. However, high-quality spectroscopy
from the ground with modern echelle spectrographs is often sufficient
to provide the means to facilitate proper model atom testing.
Figure~\ref{testing} visualises the channels available for testing a
model atom of C\,{\sc ii} using the optical spectra of early B-type stars. 
Note that despite this comprises a fair number of energy levels and
transitions from several multiplets, only a small fraction of the entire model atom can be
really scrutinised by this. Consideration of lines from additional ionization
stages (C\,{\sc iii} and C\,{\sc iv} in that case, NP08) may put further constraints, 
as the full set of observed lines should be reproduced closely by the model simultaneously. 

The above example is typical for elements with relatively simple
electron structure. More complex electron configurations like the open
3$d$-shells in the iron-group members pose a larger challenge at first
glance, but the enormous number of observable transitions puts many
constraints on the model atom construction as well. The real challenge
are therefore the low-abundance light elements lithium, beryllium and
boron, and in particular 
their alkali-like ions\footnote{High-quality atomic data 
can fortunately be obtained with relative ease for these simple ions.}. There, typically only 
the resonance lines are
observable, which gives only marginal constraints for tests of the model atom. 
Where possible, resonance lines from another ionization
stage should therefore be investigated, or subordinate lines that may
become observable in stars with particular high abundance in the
element under study. 

It is of utmost importance to use realistic atmospheric
structures for testing model atoms, requiring a proper determination of the 
stellar atmospheric parameters, see Nieva \& Przybilla (this volume).
Well-studied standard stars like the Sun (G2\,V), Procyon (F5\,IV-V), 
Vega (A0\,V), $\tau$\,Sco (B0.2\,V) or Arcturus (K1.5\,III) with tight constraints 
on their atmospheric parameters are therefore primary objects for the comparison of the models 
with observation. However, further stars that bracket the extremes of
the parameter space to be studied have also to be considered. 

\section{Final Remarks}
The plethora of possibilities implies that the model atom construction 
does not result in a unique solution. A good reproduction of
observations may be achieved by a whole family of models. The main
insight is that there are many insufficient model atoms,
but few adequate ones. In consequence, non-LTE analyses are not
superior to LTE investigations {\em per se}, but require robust and
comprehensive model atoms. In view of the increasing availability of 
accurate and precise atomic data from {\em ab-initio} calculations and 
experiments one is faced by a perpetual challenge: the impact of new 
high-quality atomic data should be tested on the modelling whenever such
become available. Of course, the same is true for new observations that
may facilitate the predictive power of the models to be tested further by
opening up other channels for the model atom calibration.

%%-----------------------------
%%      your bibliography
%%-----------------------------

\end{document}